\documentclass{osa-article}

\journal{oe}
\usepackage{siunitx}
\usepackage{todonotes}
\usepackage[capitalise]{cleveref}

\begin{document}

\title{Optimized single-shot laser ablation of concave mirror templates on optical fibers}

\author{Thibaud Ruelle,\authormark{1} Martino Poggio,\authormark{1} and Floris Braakman \authormark{1,*}}

\address{\authormark{1}Department of Physics, University of Basel, Klingelbergstrasse 82, 4056 Basel, Switzerland}

\email{\authormark{*}floris.braakman@unibas.ch} %% email address is required

% \homepage{http:...} %% author's URL, if desired

%%%%%%%%%%%%%%%%%%% abstract %%%%%%%%%%%%%%%%
%% [use \begin{abstract*}...\end{abstract*} if exempt from copyright]

\begin{abstract}
We realize mirror templates on the tips of optical fibers using a single-shot CO\textsubscript{2} laser ablation procedure.
We perform a systematic study of the influence of the pulse power, pulse duration, and laser spot size on the radius of curvature, depth, and diameter of the mirror templates.
We find that these geometrical characteristics can be tuned to a larger extent than has been previously reported, and notably observe that compound convex-concave shapes can be obtained.
This detailed investigation should help further the understanding of the physics of CO\textsubscript{2} laser ablation processes and help improve current models.
We additionally identify regimes of ablation parameters that lead to mirror templates with favorable geometries for use in cavity quantum electrodynamics and optomechanics.
\end{abstract}

%%%%%%%%%%%%%%%%%%%%%%%%%%  body  %%%%%%%%%%%%%%%%%%%%%%%%%%
\section{Introduction}

In the last decade CO\textsubscript{2} laser ablation has become a mature technique for the processing of optical glasses, uniquely suited to fabricating micrometer-scale structures with sub-nanometer surface roughness\cite{StaupendahlLasermaterialprocessing1997, WeingartenGlassprocessingpulsed2017}.
A wide range of shapes can be produced using this technique, including microspheres\cite{CollotVeryHighWhisperingGallery1993}, microlenses\cite{PaekFormationSphericalLens1975, WakakiMicrolensesmicrolensarrays1998}, microtoroids\cite{ArmaniUltrahighQtoroidmicrocavity2003}, gratings\cite{StaupendahlLasermaterialprocessing1997}, holographic structures\cite{WlodarczykDirectCO2laserbased2016} and concave mirror templates\cite{HungerLasermicrofabricationconcave2012, PetrakFeedbackcontrolledlaserfabrication2011}.
In particular, such concave mirror templates can be realized on the tip of optical fibers\cite{HungerLasermicrofabricationconcave2012}, and can be used to define tunable open-access Fabry-Perot microcavities with high finesse \cite{HungerfiberFabryPerot2010,MullerUltrahighfinesselowmodevolumeFabry2010}.
The combination of spectral tunability, high finesse, intrinsic fiber coupling and uniquely small dimensions offered by these optical cavities has led to their widespread adoption in the cavity quantum electrodynamics (CQED) community\cite{ColombeStrongatomfield2007,MullerCouplingepitaxialquantum2009,Toninelliscanningmicrocavitysitu2010,SteinerSingleIonCoupled2013,Miguel-SanchezCavityquantumelectrodynamics2013,AlbrechtCouplingSingleNitrogenVacancy2013,BrandstatterIntegratedfibermirrorion2013,BesgaPolaritonBoxesTunable2015}, and to a lesser extent in the optomechanics community\cite{FaveroFluctuatingnanomechanicalsystem2009,Flowers-JacobsFibercavitybasedoptomechanicaldevice2012,StapfnerCavityenhancedopticaldetection2013,Zhongmillikelvinallfibercavity2017}.

Over the past few years, the main focus of research on CO\textsubscript{2} laser ablation of mirror templates on optical fibers has been developing multi-shot ablation procedures in order to achieve better control over the geometry of the concave shape\cite{TakahashiNovellasermachining2014,OttMillimeterlongfiberFabryPerot2016,GarciaDualwavelengthfiberFabryPerot2018,CuiPolarizationnondegeneratefiber2018}.
These studies, alongside with improved analytical and numerical cavity models\cite{GallegoHighfinessefiberFabry2016,BenedikterTransversemodecouplingdiffraction2015,Bickrolemodematch2016,PodoliakHarnessingmodemixing2017,UphoffFrequencysplittingpolarization2015}, notably contributed to recent breakthroughs in trapped ion CQED\cite{TakahashiStrongcouplingsingle2018}, trapped atom CQED\cite{GallegoStrongPurcelleffect2018}, solid-state QED\cite{BenedikterCavityEnhancedSinglePhotonSource2017}, and optomechanics\cite{KashkanovaOptomechanicssuperfluidhelium2017}.
Nevertheless, a systematic study of the effects of fabrication parameters on the geometrical characteristics of structures created by a single ablation pulse has not yet been reported.

In this work we focus on single-shot ablation, adding to the pioneering work from \cite{HungerLasermicrofabricationconcave2012}.
We realize mirror templates on the tips of a large number of optical fibers using varying pulse powers (0.5 to \SI{3}{\watt}), pulse durations (10 to \SI{30}{\ms}), and spot sizes (32 to \SI{67}{\um}).
We characterize in detail the influence of each of those three ablation parameters on the shape of the resulting structures, and more specifically on their radius of curvature, depth, and diameter.
We then study the relationships between those three geometrical characteristics and identify regimes of ablation parameters that lead to templates with favorable geometries for use in CQED and optomechanics.

\section{Setup and methods}

\subsection{CO$_2$ ablation setup}

\begin{figure}[ht!]
\centering\includegraphics[scale=0.75]{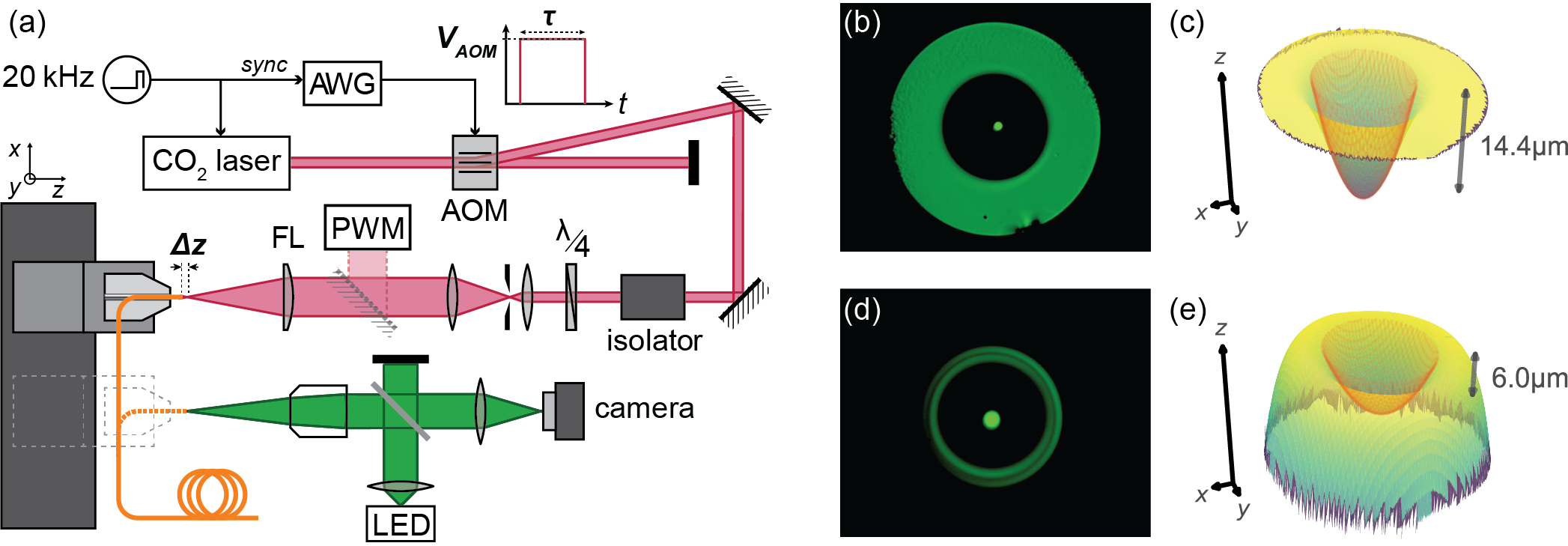}
\caption{\label{fig:setup}
(a): CO\textsubscript{2} ablation setup.
AWG: arbitrary waveform generator, AOM: acousto-optic modulator, isolator: Faraday isolator, PWM: power meter, FL: aspheric lens used to focus the CO\textsubscript{2} ablation beam, $\Delta z$: distance between the focal point of the focusing lens and the ablation target.
The inset shows the electronic pulse of amplitude $V_\mathrm{AOM}$ and duration $\tau$ that is sent to the AOM to shape the CO\textsubscript{2} ablation pulse. 
(b,d): Microscope images of two different fibers after ablation.
Only surfaces close to perpendicular to the illumination beam appear.
(c,e): Corresponding height profiles, measured by confocal laser profilometry.
The red wireframe is obtained by fitting an elliptic paraboloid to the concave structure.
The annotation shows the depth of the structure, and the coordinate system is to scale.
}
\end{figure}

Our CO\textsubscript{2} laser ablation setup is depicted in \cref{fig:setup}(a).
Similarly to the setup introduced in \cite{OttMillimeterlongfiberFabryPerot2016}, it comprises both a CO\textsubscript{2} ablation arm and an imaging arm.
The fiber holder is translated between the two arms using a Thorlabs DDS220 long-travel x stage, and is further positioned using PI M-111.12S short-travel y-z stages.
The fiber is held in a Thorlabs HFV002 V-groove fiber holder.
%not true for this series ... Here we use a custom holder which can house up to 10 

A Synrad Firestar V30 CO\textsubscript{2} laser is driven by a \SI{20}{\kilo\hertz} pulse-width-modulated control signal which ensures a continuous output and determines the maximum power that can be used for ablation.
Here we use a \SI{50}{\percent} duty cycle, which corresponds to a maximum power of \SI{2.1}{\watt}.
The laser beam is sent through a Brimrose GEM-40 acousto-optic modulator (AOM) whose RF drive is amplitude-modulated by an arbitrary waveform generator (AWG) which is synchronized to the CO\textsubscript{2} laser drive signal.
We use square pulses of amplitude $V_{\mathrm{AOM}}$ between 0 and \SI{1}{\volt} and of duration $\tau$ ranging from 10 to \SI{50}{\milli\second}.
Given that the $\SI{125}{\ns}$ rise time of the AOM is short compared to the duration of the pulses, the temporal profile imprinted on the CO\textsubscript{2} laser beam is approximately square.
The first-order diffracted beam goes through a Faraday isolator and a quarter-waveplate, ensuring the beam is circularly polarized.
The laser beam is then expanded in a Kepler telescope so that it fills the focusing lens.
A \SI{50}{\micro\meter} diameter pinhole positioned at the focal point of the beam expander provides spatial filtering.
The beam is finally focused on the target with a \SI{50}{\milli\meter} focal length aspheric lens.
The radius of the beam at the position of the target is controlled by positioning the target at a distance $\Delta z$ from the focal point of the lens, ranging from 0 to \SI{-1}{\milli\meter}.
The mapping between the defocusing distance $\Delta z$ and the $1/e^2$ beam radius (spot size) $w$ was calibrated by performing a series of knife-edge measurements of the beam profile for different values of $\Delta z$.
Additionally, a mirror can be placed between the beam expander and the focusing lens in order to measure the incident power with a power meter.
This provides a mapping between the amplitude $V_{AOM}$ of the pulse sent to the AOM and the peak power $P_{\mathrm{CO2}}$ of the ablation pulse.

The imaging arm consists of an infinity-corrected microscope objective, a green LED and a CMOS camera.
The green light from the LED is collimated and directed to the objective with a beamsplitter.
The light reflected from the target is then focused on the camera sensor by a tube lens.

\subsection{Experimental procedure}

We use bare Thorlabs 780HP single mode fiber, which has a cladding diameter of \SI{125}{\um} and a core diameter of \SI{8}{\um}.
Each fiber is prepared, ablated, and finally characterized.

% Fiber preparation
The fiber preparation steps include cutting to length, stripping, cleaving, and positioning into the holder.
We use a Photon Kinetics PK11 ultrasonic cleaver to obtain smooth cleaves with minimal cleave angles.
The fiber is clamped onto the V-groove holder with about \SI{2}{\mm} of freestanding length. 

% ablation procedure
Before ablation, the core of the fiber is centered in front of the imaging arm.
The holder is then translated by a carefully calibrated distance in all three directions so that the core of the fiber is located at the focal point of the laser beam.
An additional displacement $\Delta z$ (defocusing distance) along the beam axis can be performed at this point in order to change the spot size of the CO\textsubscript{2} beam at the position of the fiber.
We then signal the AWG to send an electrical pulse of amplitude $V_\mathrm{AOM}$ and duration $\tau$ to the AOM at the next raising edge of the CO\textsubscript{2} laser control signal.
After ablation, the target is translated back to its original position for imaging.
In-situ microscopy (\cref{fig:setup}(b) and (d)) is used to confirm centering and to roughly estimate the geometrical characteristics of the concave structure.
The fiber is finally transferred to a different holder for profiling and storage.

% imaging
For characterization, each fiber is profiled with a Keyence VK-X200K laser scanning confocal microscope.
The two examples of such profiles plotted in \cref{fig:setup}(c) and (e) show two typical geometries that we obtain depending on the ablation pulse power, as discussed in the next section.
We additionally performed AFM measurements on a few of the fibers and found that the roughness at the center of the ablated structures is typically smaller than \SI{0.3}{\nm} rms, in good agreement with previously reported values\cite{HungerLasermicrofabricationconcave2012}.

\subsection{Fitting procedure}

\begin{figure}[ht!]
\centering\includegraphics[scale=0.75]{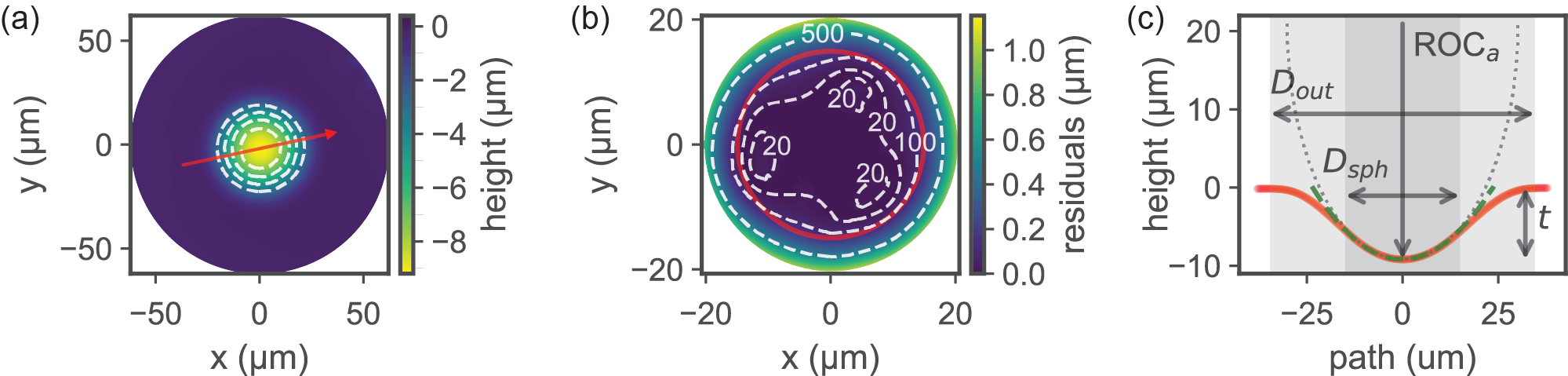}
\caption{\label{fig:fitting}
Graphical output of the fitting routine.
(a): Height profile of a fiber as measured by laser scanning confocal microscopy.
The white dashed lines are contour lines at -7.6, -6.1, -4.5, and \SI{-2.9}{\nm} starting from the center.
The red arrow shows the path of the linecut plotted in (c).
(b): Residuals of the fit of an elliptic paraboloid to the height profile in (a), with white dashed contour lines.
The red circle shows the disk of diameter $D_\mathrm{sph}$ within which we consider the structure to be equivalent to a sphere of radius ROC.
(c): Linecut through the height profile data in (a), taken along the major axis of the best fit elliptic paraboloid.
The green dashed curve is the corresponding linecut through the best fit elliptic paraboloid.
The dotted black curve is the circle of radius ROC$_a$ (defined in the main text), which appears as an ellipse due to the scale used.
The annotations illustrate the geometrical characteristics that are calculated during the fitting procedure.
}
\end{figure}

When the height profiles exhibit a concave structure, we run them through a Python fitting routine that extracts the characteristic dimensions of the concave part.
The fitting process is illustrated in \cref{fig:fitting}, in which the graphical output of the routine is displayed.
The Python script first corrects for plane tilt, finds the center of the structure, and calculates its outer diameter $D_\mathrm{out}$ which we define as the diameter of the contour line at \SI{5}{\percent} of the depth of the structure.
An elliptic paraboloid is then fitted to the height profile cropped to a disk centered on the structure and whose radius is half the waist of a Gaussian fitted to a linecut through the structure.
The fit results are then used to calculate the depth $t$ of the structure and its radii of curvature $\mathrm{ROC}_a$ and $\mathrm{ROC}_b$ along the major and minor axes of the elliptic paraboloid.
We additionally define the mean radius of curvature $\mathrm{ROC} = (\mathrm{ROC}_a + \mathrm{ROC}_b)/2$, and the asymmetry $\gamma = (\mathrm{ROC}_a - \mathrm{ROC}_b)/\mathrm{ROC}_a$.
The routine finally computes the residuals of the fit and calculates the spherical diameter $D_\mathrm{sph}$ of the structure, which we define as the diameter of the disk centered on the structure for which fit residuals are smaller than \SI{100}{\nano\meter}.
$D_\mathrm{sph}$ is intended to be an estimate of the effective mirror diameter as used in \cite{HungerfiberFabryPerot2010}.

\section{Effects of ablation parameters on the shape of the structure}

CO\textsubscript{2} laser ablation is a complex, multi-physical process in which the dynamics of heat transfer, phase transitions and liquid flow all come into play.
The dominant phenomena for determining the shape created at the ablation site are strongly material dependent and change for even small variations in ablation parameters\cite{DoualleThermomechanicalsimulationsCO22016}.
For instance, when the surface temperature is not raised above the vaporization temperature, material removal is minimal and the ablation site mainly undergoes smoothing\cite{MendezLocalizedCO2laser2006}.
When vaporization occurs, for certain ablation parameters a combination of vaporization and of melt displacement driven by recoil pressure can result in the formation of a concave shape\cite{MarkillieEffectvaporizationmelt2002}.
Finally, if solidification occurs slowly enough, capillary forces can make the geometry evolve further, eventually leading to a convex shape\cite{HeNumericalmodelexperimental2018}.
Mainly due to a lack of fundamental understanding of the ablation process and to the lack of data on material properties at high temperature, both analytical and numerical models have yet to demonstrate the ability to accurately predict the shape resulting from CO\textsubscript{2} laser ablation within an experimentally relevant range of ablation parameters\cite{FeitMechanismsCO2laser2003,NowakAnalyticalmodelCO22015,DoualleThermomechanicalsimulationsCO22016,HeNumericalmodelexperimental2018}.
Modeling is especially problematic in the case when optical fibers are used as the target because lateral boundary effects come into play\cite{HungerLasermicrofabricationconcave2012} .
As a result, the extensive experimental exploration of the fabrication parameter space presented here is relevant both from a fundamental point of view and to determine ablation parameters useful to create shapes designed for specific applications.
Extending on the work in \cite{HungerLasermicrofabricationconcave2012}, we aim at improving the understanding of the effect of each ablation parameter, extending the range of geometries that can be achieved and providing guidelines for the fabrication of structures for open micro-cavities.

We study the geometrical characteristics of 129 structures machined on Thorlabs 780HP single mode fibers following the single-shot CO\textsubscript{2} ablation procedure described above.
We used pulse durations $\tau$ of 10, 30, and \SI{50}{\milli\second} and defocusing distances $\Delta z$ of 0, -0.1, -0.2, -0.3, and \SI{-0.4}{\milli\meter}, corresponding to spot sizes $w$ of 32, 36, 45, 56, and \SI{67}{\um}.
For each combination of those parameters we performed a series of ablations with varying pulse power $P_{\mathrm{CO2}}$.
We then measured the height profile of the fiber facets, which can be flat, concave, convex or a mixture of convex and concave.
We focus on the 123 fibers that are concave in their centers and fit this concave part to extract the geometrical characteristics of the structure.

\begin{figure}[ht!]
\centering\includegraphics[scale=0.75]{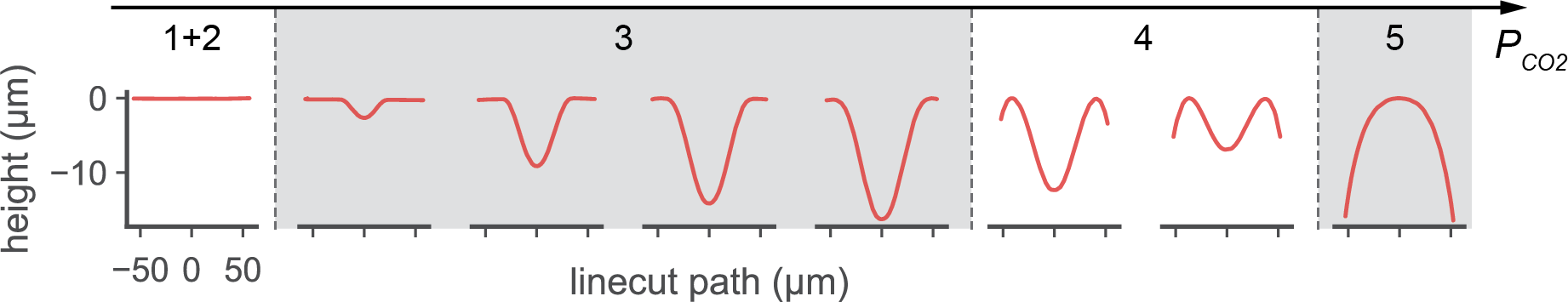}
\caption{\label{fig:linecuts}
Linecuts taken through the height profiles of fibers ablated with increasing ablation pulse power and duration, as measured by laser scanning confocal microscopy.
The axis shows the direction along which $P_\mathrm{CO2}$ increases and illustrates the power regimes defined in the main text.
A spot size of \SI{45}{\micro\meter} was used for all fibers.
A pulse duration of \SI{30}{\ms} was used for all fibers but the rightmost, which was subjected to a \SI{300}{\ms} ablation pulse.
}
\end{figure}

In order to illustrate the effect of ablation pulse power on the shape, linecuts through the height profiles of a selection of fibers are plotted in \cref{fig:linecuts}.
We distinguish 5 different regimes of pulse power, each of them leading to a different type of modification of the surface of the fiber, some of which might not be observed depending on the value of the other ablation parameters.
1) For very low pulse powers, no modification of the surface occurs.
2) For low pulse powers, the overall geometry is not modified, but the area exposed to the laser is smoothed due to melting and resolidification.
3) For medium pulse powers, concave structures are created, whose depth and outer diameter increase with pulse power.
4) For high pulse powers, concave structures whose depth and outer diameter decrease with pulse power are created within an increasingly convex shape.
This compound shape is an interesting geometry for short cavities \cite{KauppPurcellEnhancedSinglePhotonEmission2016}.
5) For very high pulse power, a fully convex shape is created. 
We are mainly interested in regimes 3 and 4, in which a concave structure is created.
These two regimes must be considered separately for further discussion of the effects of pulse power, pulse duration, and spot size on the geometry of the structure.

\begin{figure}[ht!]
\centering\includegraphics[scale=0.75]{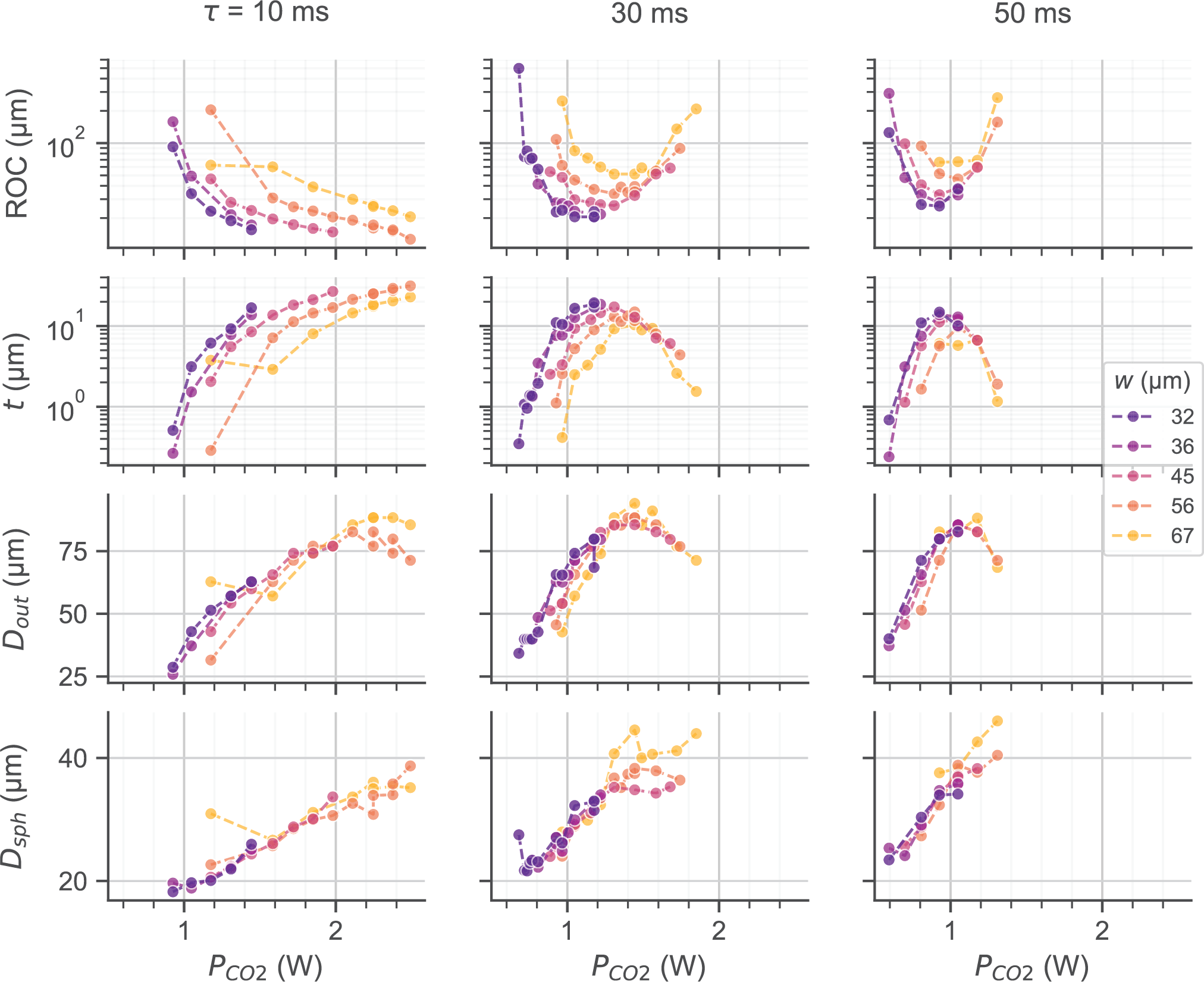}
\caption{\label{fig:cratersVsCO2}
Geometrical characteristics of ablated concave structures plotted as a function of ablation parameters.
The different rows show plots of $\mathrm{ROC}$, $t$, $D_\mathrm{out}$, and $D_\mathrm{sph}$ as a function of the pulse power $P_{\mathrm{CO2}}$.
The pulse duration $\tau$ is varied across columns, and the spot size $w$ is encoded in the color of the points.
}
\end{figure}

The geometrical characteristics $\mathrm{ROC}$, $t$, $D_\mathrm{out}$, and $D_\mathrm{sph}$ obtained by fitting the profiles of the fibers exhibiting concave structures, are plotted in \cref{fig:cratersVsCO2} as a function of the ablation parameters $P_{\mathrm{CO2}}$, $\tau$, and $w$.
In the low power regime (regime 3), an increase in pulse power leads to a decrease in ROC, and an increase in depth, outer diameter and spherical diameter.
In the high power regime (regime 4), an increase in pulse power leads to an increase in ROC, a decrease in depth and outer diameter, and an increase in spherical diameter.
In both power regimes, an increase in spot size leads to an increase in ROC, a decrease in depth and has no significant effect on the outer and spherical diameter.
Increasing the pulse duration has a more complex effect: it generally shifts values of the geometrical characteristics to lower pulse powers and narrows their distribution.
This results in a decrease in the pulse power corresponding to the onset of regime 4, associated to an increase in the sensitivity of the geometrical characteristics to changes in the pulse power.
As a consequence, shorter pulse durations give a finer control over the geometry of the structures since deviations in pulse power have a smaller effect.
In addition, decreasing the pulse duration decreases the minimum achievable ROC, increases the maximum achievable depth, and decreases the minimum achievable outer and spherical diameter.
Note that we do not observe the defocusing distance to have a significant effect on crater asymmetry, which we measure to be \SI{5}{\percent} on average.

\section{Relationships between the geometrical characteristics of concave structures}

Fiber-based Fabry-Perot optical microcavities are widely used in the fields of CQED \cite{ColombeStrongatomfield2007,MullerCouplingepitaxialquantum2009,Toninelliscanningmicrocavitysitu2010,SteinerSingleIonCoupled2013,Miguel-SanchezCavityquantumelectrodynamics2013,AlbrechtCouplingSingleNitrogenVacancy2013,BrandstatterIntegratedfibermirrorion2013,BesgaPolaritonBoxesTunable2015,GallegoStrongPurcelleffect2018,TakahashiStrongcouplingsingle2018} and optomechanics\cite{FaveroFluctuatingnanomechanicalsystem2009,Flowers-JacobsFibercavitybasedoptomechanicaldevice2012,StapfnerCavityenhancedopticaldetection2013,Zhongmillikelvinallfibercavity2017}, with additional applications in sensing \cite{Maderscanningcavitymicroscope2015,PetrakPurcellenhancedRamanscattering2014,HummerCavityenhancedRamanmicroscopy2016}.
A figure of merit common to most of these applications is the ratio between the finesse $\mathcal{F}$ of the cavity and the cross-section $\pi w_0^2$ of the fundamental mode of the cavity, which we define here as $\,\mathrm{FOM}=\mathcal{F}/(\pi w_0^2)$.
In order to maximize finesse and minimize the waist $w_0$, it is necessary to optimize several geometrical characteristics simultaneously, while complying with experimental requirements specific to each application.
However, strong relationships exist between the different geometrical characteristics of concave structures created by CO\textsubscript{2} ablation, which have been reported to put most of the parameter space out of reach \cite{Greutersmallmodevolume2014,NajerFabricationmirrortemplates2017}, leading to significant compromises.
In this section we study the relationships between the geometrical characteristics of the concave structures, showing that they can be independently chosen to a larger extent by varying $\tau$, $\Delta z$, and $P_\mathrm{CO2}$.
We then point towards strategies to fabricate mirror templates tailored for two commonly used cavity geometries and their associated applications.

\begin{figure}[ht!]
\centering\includegraphics[scale=0.75]{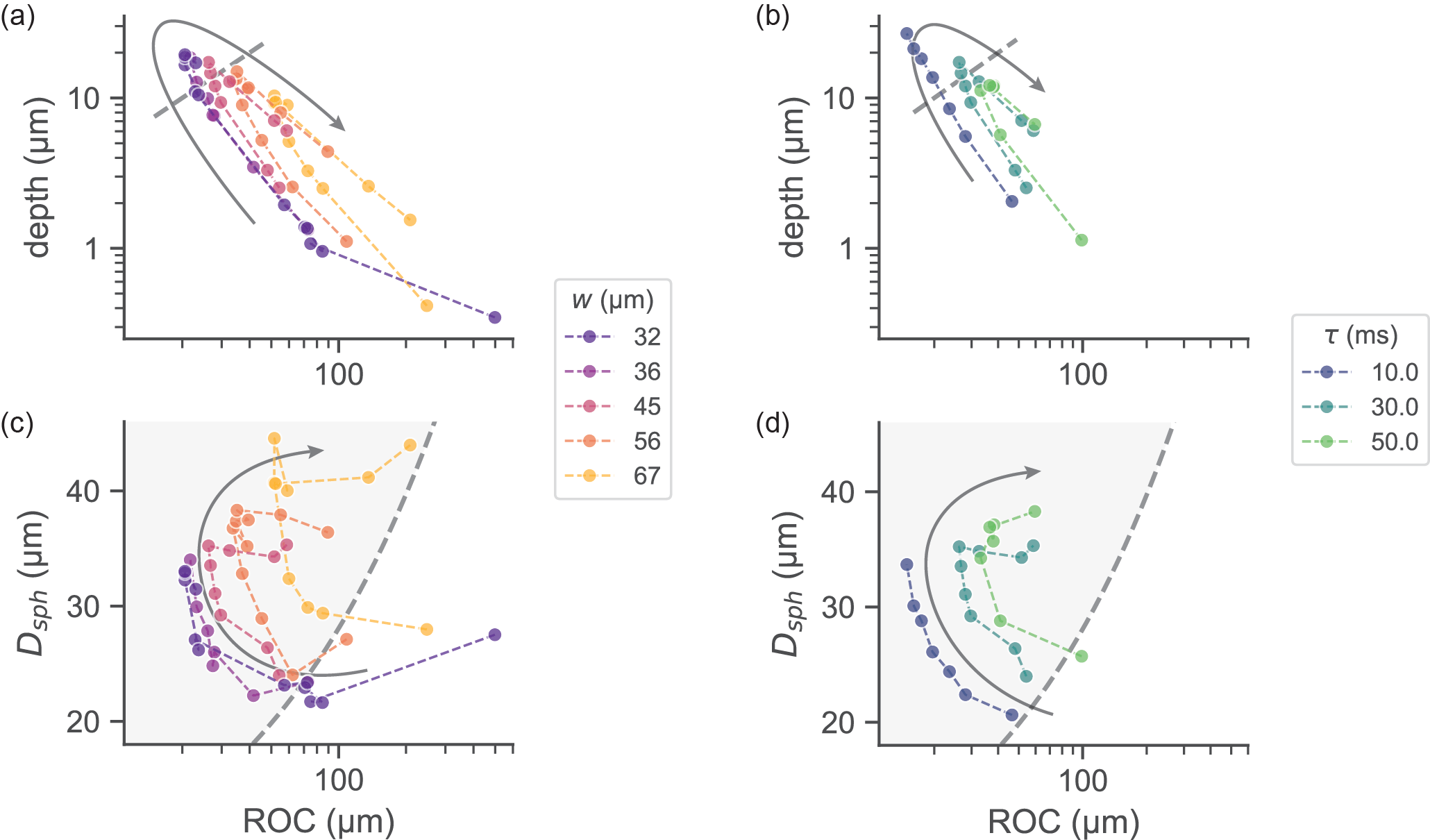}
\caption{\label{fig:correlations}
Plots of the relationship between selected geometrical characteristics as a function of selected ablation parameters.
(a): Structure depth as a function of radius of curvature for various spot sizes and for a pulse duration of \SI{30}{\ms}.
(b): Structure depth as a function of radius of curvature for various pulse durations and for a spot size of \SI{45}{\um}.
(c): Spherical diameter as a function of radius of curvature for various spot sizes and for a pulse duration of \SI{30}{\ms}.
(d): Spherical diameter as a function of radius of curvature for various pulse durations and for a spot size of \SI{45}{\um}.
The gray arrows show the direction of increasing pulse power.
In (a,b) the dashed line follows $t=\mathrm{ROC}/2$.
In (c,d) the shaded area shows the region where clipping losses are small, as defined by \cref{eq:Dsph}.
}
\end{figure}

The relationship between the radius of curvature and the depth of the structures is shown in \cref{fig:correlations}(a,b).
It is most relevant to fiber-based cavity QED with solid-state emitters, or to other applications where an optical emitter is located on or near one of the mirrors.
The optimal cavity geometry is the planar-concave geometry, for which the waist of the fundamental mode is given by:
\begin{equation}
	w_0^2 = \frac{\lambda L_\mathrm{cav}}{\pi}\sqrt{\frac{1}{\varepsilon} - 1}\ , \quad \text{with} \quad \varepsilon = \frac{L_\mathrm{cav}}{\mathrm{ROC}}\ ,\  0 \leq \varepsilon \leq 1\ ,
\end{equation}
where $\mathrm{ROC}$ is the radius of curvature of the concave mirror and $L_\mathrm{cav}$ is the cavity length.
FOM is usually improved by minimizing both the cavity length and the radius of curvature in order to decrease the mode waist, the physical limit for $L_\mathrm{cav}$ being the depth of the structure.
One should additionally make sure that $L_\mathrm{cav} < \mathrm{ROC}/2$ in order to prevent finesse deterioration due to clipping losses\cite{BenedikterTransversemodecouplingdiffraction2015}.
A guideline for the best cavity geometry is then $L_\mathrm{cav} = t = \mathrm{ROC}/2$ (shown as a gray dashed line in \cref{fig:correlations}(a,b)), with $\mathrm{ROC}$ and $t$ as small as possible.
The ablation results plotted in \cref{fig:correlations}(a,b) show a strong nonlinear relationship between the radius of curvature and the depth of structures ablated with varying pulse power at constant spot size and pulse duration.
However we observe that this relationship depends strongly on the value of the spot size and pulse duration.
Structures with favorable geometries can be produced using short pulse durations and small spot sizes, the latter of which can be obtained by using a focusing lens with a larger numerical aperture and by performing the ablation at its focal point.

In contrast, another category of applications exists for which experimental constraints limit how short the cavity can be made.
Fiber-based cavity QED with trapped atoms or ions, fiber-based cavity optomechanics, or other applications where the emitter or mechanical resonator is located in between the two mirrors belong to this category.
The preferred cavity geometry is then the symmetric geometry, and for these relatively long cavity lengths the main obstacle to maximizing FOM is maintaining a high finesse.
Finesse is degraded by clipping losses, which arise when the spot size of the fundamental mode of the cavity on the end mirrors becomes large compared to the spherical diameter.
The condition on $D_\mathrm{sph}$ for clipping losses not to significantly degrade the finesse of a symmetric cavity is given by\cite{HungerfiberFabryPerot2010}:
\begin{equation}\label{eq:Dsph}
D_\mathrm{sph}^2 \geq \ln \left(\frac{5\mathcal{F}}{\pi}\right)\frac{\lambda L_\mathrm{cav}}{\pi\sqrt{\varepsilon\,(2-\varepsilon)}}\ , \quad \text{with} \quad \varepsilon = \frac{L_\mathrm{cav}}{\mathrm{ROC}}\ ,\  0 \leq \varepsilon \leq 2\ .
\end{equation}
In order to minimize waist while maintaining a high finesse, one should choose structures with the smallest radii possible which satisfy both \cref{eq:Dsph} and $\mathrm{ROC} > L_\mathrm{cav}$.
The relationship between the spherical diameter and the radius of curvature of the structures is shown in \cref{fig:correlations}(c,d), where the region defined by \cref{eq:Dsph} is shown for $L_\mathrm{cav} = \mathrm{ROC}$.
Craters fabricated with a high pulse power within regime 4 exhibit the largest spherical diameters, while pulse duration or spot size can be changed to adjust the radius of curvature.

\section{Conclusions}
In conclusion, we have performed a systematic study of the effect of single-shot CO\textsubscript{2} laser ablation parameters on fiber tip geometry. 
We have investigated the effects of extended ranges of pulse power, pulse duration, and spot size and developed guidelines for the fabrication of fiber mirror templates optimized for experiments in optomechanics and CQED. 
We anticipate that this study will be helpful in providing empirical insight in the physics of laser ablation processes.
Notably, we have observed a new parameter regime in which $\mathrm{ROC}$, $D_\mathrm{sph}$, and $D_{\mathrm{out}}$ exhibit extrema as a function of $P_{\mathrm{CO2}}$ and $\tau$ (see \cref{fig:cratersVsCO2}). 
We speculate that the presence of such extrema arises from the interplay of surface tension and boundary effects originating from the limited lateral size of the fibers. 
Associated with this we find that compound concave-convex shapes can be produced using a simple single-shot ablation procedure (ablation regime 4 in \cref{fig:linecuts}). 
Such compound shapes are very useful for small mode volume cavities and previously required tapering of the fiber in an additional processing step\cite{KauppPurcellEnhancedSinglePhotonEmission2016}.
Finally, we expect that using shorter laser pulses and smaller spot sizes than presented here will enable to simultaneously decrease $\mathrm{ROC}$ and crater depth further.
%It is relevant to a new type of optomechanical devices that is currently gaining traction: suspended membranes made of few-layer two-dimensional materials\cite{NortheastSuspensionsimpleoptical2018,BartonPhotothermalSelfOscillationLaser2012}.
%Such membranes are typically stamped on substrates with pre-patterned holes, thus defining free-standing suspended drum resonators which resonance frequency is determined by the outer shape of the holes \cite{Castellanos-GomezMechanicsfreelysuspendedultrathin2015}.
\section*{Funding}
%Please identify all appropriate funding sources by name and contract number. Funding information should be listed in a separate block preceding any acknowledgments. List only the funding agencies and any associated grants or project numbers, as shown in the example below:\\
%\\
%National Science Foundation (NSF) (1253236, 0868895, 1222301); Program 973 (2014AA014402); Natural National Science Foundation (NSFC) (123456).\\
%\\
%OSA participates in \href{https://www.crossref.org/fundingdata/}{Crossref's Funding Data}, a service that provides a standard way to report funding sources for published scholarly research. To ensure consistency, please enter any funding agencies and contract numbers from the Funding section in Prism during submission or revisions.
We acknowledge the support of the ERC through Starting Grant NWScan (Grant No. 334767), the Swiss National Science Foundation (Ambizione Grant No. PZ00P2-161284/1) and the NCCR Quantum Science and Technology (QSIT).

\section*{Acknowledgments}
%Acknowledgments, if included, should appear at the end of the document. The section title should not be numbered.
We thank Lukas Greuter, Daniel Najer, Daniel Riedel, and Richard Warburton for assistance with the ablation setup and helpful discussions. Furthermore, we thank Sascha Martin and the mechanical workshop at the University of Basel Department of Physics for technical support.

%
%%%%%%%%%%%%%%%%%%%%%%%% References %%%%%%%%%%%%%%%%%%%%%%%%%
%
%Add references with BibTeX or manually.
%\cite{Zhang:14,OSA,FORSTER2007,Dean2006}
%
%%%%%%%%%%% If using BibTeX:
\bibliography{TR-201808-CO2_ablation_paper}

%%%%%%%%%% If preparing manually:
% \begin{thebibliography}{1}
% \newcommand{\enquote}[1]{``#1''}

% \bibitem{Zhang:14}
% Y.~Zhang, S.~Qiao, L.~Sun, Q.~W. Shi, W.~Huang, L.~Li, and Z.~Yang,
%   \enquote{Photoinduced active terahertz metamaterials with nanostructured
%   vanadium dioxide film deposited by sol-gel method,}
%   {\protect\JournalTitle{Optics Express}} \textbf{22}, 11070--11078 (2014).

% \bibitem{OSA}
% {Optical Society}, \enquote{{OSA Publishing},}
%   \url{http://www.osapublishing.org}.

% \bibitem{FORSTER2007}
% P.~Forster, V.~Ramaswamy, P.~Artaxo, T.~Bernsten, R.~Betts, D.~Fahey,
%   J.~Haywood, J.~Lean, D.~Lowe, G.~Myhre, J.~Nganga, R.~Prinn, G.~Raga,
%   M.~Schulz, and R.~V. Dorland, \enquote{Changes in atmospheric consituents and
%   in radiative forcing,} in \enquote{Climate Change 2007: The Physical Science
%   Basis. Contribution of Working Group 1 to the Fourth assesment report of
%   Intergovernmental Panel on Climate Change,}  S.~Solomon, D.~Qin, M.~Manning,
%   Z.~Chen, M.~Marquis, K.~B. Averyt, M.~Tignor, and H.~L. Miler, eds.
%   (Cambridge University Press, 2007).

% \end{thebibliography}

\end{document}